# Constrained realizations and minimum variance reconstruction of non-Gaussian random fields

Ravi K. Sheth
*Berkeley Astronomy Department, University of California, Berkeley, CA 94720*

**ABSTRACT**
With appropriate modifications, the Hoffman–Ribak algorithm that constructs constrained realizations of Gaussian random fields having the correct ensemble properties can also be used to construct constrained realizations of those non-Gaussian random fields that are obtained by transformations of an underlying Gaussian field. For example, constrained realizations of lognormal, generalized Rayleigh, and chi-squared fields having $n$ degrees of freedom constructed this way will have the correct ensemble properties. The lognormal field is considered in detail.

For reconstructing Gaussian random fields, constrained realization techniques are similar to reconstructions obtained using minimum variance techniques. A comparison of this constrained realization approach with minimum variance, Wiener filter reconstruction techniques, in the context of lognormal random fields, is also included. The resulting prescriptions for constructing constrained realizations as well as minimum variance reconstructions of lognormal random fields are useful for reconstructing masked regions in galaxy catalogues on smaller scales than previously possible, for assessing the statistical significance of small-scale features in the microwave background radiation, and for generating certain non-Gaussian initial conditions for $N$-body simulations.

## 1 INTRODUCTION

The recovery of a signal from noisy and incomplete data is a classic inversion problem that is of particular relevance when analysing observations of the galaxy distribution for two reasons. First, regions like the 'zone of avoidance' that is a result of obscuration by the Galactic Plane prohibit the construction of complete all-sky catalogues. The existence of 'masked' regions raises intriguing questions about the connectivity of structures separated by a masked region, and so about the scale of the largest coherent structures in the observed Universe. Secondly, if one assumes that the luminous galaxies that are observed sample an underlying density field that is smooth, then the discreteness of objects introduces 'shot-noise' that may inhibit the reconstruction of the underlying smooth field from the data.

Recently, Lahav et al. (1994) addressed these problems of incomplete sky coverage and shot-noise within the context of whole-sky galaxy surveys. They note that, when the underlying density field is a Gaussian random field, minimum variance techniques (e.g. Rybicki & Press 1992) and the conditional probability framework of constrained realizations (Hoffman & Ribak 1991) both provide the same estimate for the 'optimal reconstruction' of the underlying field, given the observed field. This is essentially a consequence of three properties peculiar to Gaussian random fields. First, a Gaussian field is completely specified by two parameters, essentially its mean and variance, so that minimizing the variance between an observed Gaussian field and the reconstructed Gaussian field will certainly produce an optimal estimate of the underlying field. Secondly, for Gaussian fields, the most probable field is also the mean field, so that reconstruction of the mean field, given the observed field, also gives the optimal estimate of the true underlying field. Thirdly, the distribution of residuals around the mean realization of a Gaussian random field is also Gaussian, with a variance that is independent of the value of the mean realization. As Hoffman & Ribak (1991) showed, with the previous two properties, this third property admits a powerful, efficient technique for constructing constrained realizations (e.g. Bertschinger 1987) of Gaussian random fields.

Counts-in-cells analyses of the angular distribution of *IRAS* galaxies show that, on large angular scales (angular separations $\theta \gtrsim 10°$), the *IRAS* distribution is essentially Gaussian (e.g. Sheth, Mo & Saslaw 1994). Therefore, when Lahav et al. (1994) apply their minimum variance technique and construct constrained realization reconstructions of the angular distribution of *IRAS* galaxies, to recover structure on scales $\gtrsim 15°$, their results are spectacular. On smaller scales, however, the *IRAS* distribution is extremely non-Gaussian. Unfortunately, when the underlying distribution is non-Gaussian, the mean field reconstruction and the most probable reconstruction are not, generally, the same. It may be that reconstructions that minimize variance will differ significantly from constrained realization techniques in this regime.

Sheth et al. (1994) show that, on nonlinear scales (corresponding to small angular separations), the data appear consistent with the hypothesis that the *IRAS* galaxies are drawn from an underlying smooth density field that is a lognormal. Bernardeau & Kofman (1995) argue that there is good reason to expect a lognormal to provide a good approximation to the true density field traced by galaxies, if the initial power spectrum was similar to that which is expected in the standard cold dark matter model. Therefore, to extend the Lahav et al. (1994) analysis to smaller, nonlinear scales, a lognormal model for the non-Gaussian density field is likely to be both useful and plausible. This is because, for a lognormal distribution, it is possible to compute the difference between the minimum variance estimator of the underlying field, the mean field, and the most probable field, given the observed field. A systematic study of the differences between these estimators of the true underlying lognormal field will allow optimal whole-sky reconstructions of the galaxy density distributions even on the very small scales where the galaxy distribution is highly nonlinear. Moreover, it will help quantify the strengths, and also the limitations, of the minimum variance and the constrained realization approaches to reconstructing an underlying density field.

In the cosmological context, the ability to construct constrained realizations of random fields is also useful because it allows the generation of initial conditions for numerical simulations that are designed to incorporate, a priori, particular features of interest. For instance, rare density peaks of a Gaussian random field may behave quite differently from those in non-Gaussian fields. The study of the detailed dynamics of the collapse of density peaks is greatly facilitated by the ability to 'make them to order' in the initial conditions of a numerical simulation. Moreover, the problem of 'cosmic variance' due to the long-wave modes that are missing in a finite-sized simulation box may be partially alleviated by using those long-wave modes measured observationally to constrain the initial conditions in the simulation.

Section 2.1 summarizes useful background material about Gaussian and lognormal random fields. The notation follows that of Coles & Jones (1991). Section 2.2 computes the conditional lognormal distribution; it is a 'shifted' lognormal. Then it computes the distribution function of the residuals of a lognormal field, given a set of constraints that, say, specify the value of the field in certain regions (so the distribution of the constraints is also lognormal). Since, for lognormal fields, the mean and most-probable fields are different, the distribution of residuals from both fields is calculated. For both cases, the distribution of the residual field depends on the values of the constraints. In this respect, lognormal fields are different from Gaussian fields.

It would seem, therefore, that the constrained realization algorithm suggested by Hoffman & Ribak (1991) cannot be applied to lognormal density fields. Finally, Section 2.3 demonstrates that the Hoffman–Ribak algorithm can be used to construct constrained realizations of the log of a lognormal density field. It also shows that taking the exponential of the constrained realization of the log of the lognormal field gives a density field that is the 'optimal' constrained realization. That is, when used in this way, the Hoffman–Ribak algorithm can construct an ensemble of constrained realizations of lognormal fields that has the correct properties, just as it can for Gaussian fields. Section 3 provides an algorithm that produces constrained realizations of lognormal fields that can be compared directly with, e.g., the *IRAS* catalogue, while Section 4 relates all these results to the use of minimum variance, Wiener Filter reconstructions of lognormal random fields. This last is of interest given the demonstration (Rybicki & Press 1992; Lahav et al. 1994) of the equivalence of the constrained realization and minimum variance approaches when reconstructing Gaussian random fields.

Section 5 discusses extensions of the Hoffman–Ribak algorithm to treat other random fields that are obtained by simple transformations of an underlying Gaussian field. It argues that, analogous to the lognormal case, an appropriately modified application of the Hoffman–Ribak algorithm will also produce constrained realizations of these other non-Gaussian random fields, and that these constrained realizations will possess the correct ensemble properties. As specific examples, it shows this to be true for chi-squared fields having $n$-degrees of freedom, and for generalizations of the Rayleigh and Maxwell distributions considered by Coles & Barrow (1987).

## 2  CONSTRAINED REALIZATIONS OF LOGNORMAL RANDOM FIELDS

### 2.1  Notation

The one-point probability distribution function (pdf), $f_1(x)$, for a Gaussian random field, $X(\boldsymbol{r})$, is

$$f_1(x)\,\mathrm{d}x = \frac{\mathrm{d}x}{\sqrt{2\pi\sigma^2}}\,\exp\left[-\frac{(x-\mu)^2}{2\sigma^2}\right], \qquad (1)$$

where $\mu$ and $\sigma$ are, respectively, the mean and the variance of $X$. For a Gaussian random field all the higher order $n$-point pdf's, $f_n(\boldsymbol{x})$, of field values at different positions $\boldsymbol{r}_i$, are multivariate Gaussians so that

$$f_n(\boldsymbol{x}) = (2\pi)^{-\frac{n}{2}}|\mathbf{M}|^{-\frac{1}{2}}\exp\left[-\frac{1}{2}\sum_{i,j}M_{ij}^{-1}(x_i-\mu)(x_j-\mu)\right], \quad (2)$$

where $\boldsymbol{x}=(x_1,\ldots,x_n)$, $x_i = x(\boldsymbol{r}_i)$, and $\mathbf{M}$ is the covariance matrix with elements

$$M_{ij} = \langle (X_i-\mu)(X_j-\mu)\rangle, \qquad (3)$$

where all the $\mu_i$'s are assumed to have the same value, $\mu$.

The lognormal field (LN) is obtained simply by transforming a Gaussian field via

$$Y(\boldsymbol{r}) = \exp\left[X(\boldsymbol{r})\right], \qquad (4)$$

so that its one-point pdf is

$$f_1(y)\,\mathrm{d}y = \frac{\mathrm{d}y}{y}\,\frac{1}{\sqrt{2\pi\sigma^2}}\,\exp\left[-\frac{(\ln y-\mu)^2}{2\sigma^2}\right], \qquad (5)$$

where $\mu$ and $\sigma^2$ are the mean and variance of the underlying Gaussian field $X$. The corresponding $n$-point pdf is

$$f_n(y_1,\ldots,y_n) = (2\pi)^{-n/2}|\mathbf{M}|^{-1/2} \times \prod_{i=1}^{n}\frac{1}{y_i} \times \exp\left[-\frac{1}{2}\sum_{i,j}M_{ij}^{-1}(\ln y_i - \mu)(\ln y_j - \mu)\right], \quad (6)$$

where $\mathbf{M}$ is the covariance matrix of the $X$ values defined above.

### 2.2 Properties of the constrained field

This subsection uses the notation of the simple example of Section 2 in the Lahav et al. (1994) Wiener filter letter. The simple case they considered demonstrates all the interesting properties without having to introduce more complicated notation for a random field rather than a random variable.

First, define the following statements:
$P(x)=$ the probability the random variable has value $x$,
$P(x,y) =$ the joint probability that one variable has value $x$ and the other is $y$,
$P(x|y) =$ the probability that one variable has the value $x$ given that the other is $y$.

To construct constrained realizations, think of $x$ as the value of a realization given that the constraint has value $y$. Next, calculate the mean value of an ensemble of realizations of $x$, in which each realization of $x$ is subject to the (same) constraint $y$, $\langle x|y\rangle$, and then calculate the distribution of the residual fluctuations around this value of the mean realization, i.e., the distribution of $u \equiv x - \langle x|y\rangle$. The key to the Hoffman–Ribak algorithm is that, for a Gaussian random field, the distribution of $u$ is independent of the value of $y$. Is this true for a lognormal?

For clarity (and so that factors of $\mu$ don't proliferate) the following considers the variable $x$ which is assumed to be distributed lognormally and for which the mean, $\mu$, of the underlying Gaussian variable is zero. Then, equation (5) shows that

$$P_{\rm LN}(x) = \frac{1}{x}\frac{1}{\sqrt{2\pi\sigma_x^2}}\exp\left[-\frac{(\ln x)^2}{2\sigma_x^2}\right], \quad (7)$$

where $\sigma_x^2 = \langle(\ln x)^2\rangle$. When $\mu = 0$, equation (6) shows that the joint probability

$$P_{\rm LN}(x,y) = \frac{|\mathbf{M}|^{-\frac{1}{2}}}{2\pi xy}\exp-\frac{1}{2}\left[M_{11}^{-1}(\ln x)^2 + 2M_{12}^{-1}(\ln x\,\ln y) + M_{22}^{-1}(\ln y)^2\right], \quad (8)$$

where $\mathbf{M}$ is the covariance matrix:

$$\mathbf{M} = \begin{pmatrix} \sigma_x^2 & \xi \\ \xi & \sigma_y^2 \end{pmatrix}, \quad (9)$$

with $\sigma_x^2 = \langle(\ln x)^2\rangle$, $\sigma_y^2 = \langle(\ln y)^2\rangle$, and $\xi = \langle\ln x\,\ln y\rangle$. So, $\xi$ involves the cross-correlations between $\ln x$ and $\ln y$, the determinant $|\mathbf{M}|$ is just $\sigma_x^2\sigma_y^2 - \xi^2$, and the inverse, $\mathbf{M}^{-1}$, is easily computed. Then, the conditional probability is

$$\begin{aligned}P_{\rm LN}(x|y) &= \frac{P_{\rm LN}(x,y)}{P_{\rm LN}(y)}\\ &= \frac{1}{x}\frac{1}{\sqrt{2\pi}}\sqrt{\frac{\sigma_y^2}{\sigma_x^2\sigma_y^2 - \xi^2}}\\ &\quad \times \exp\left[-\frac{1}{2}\left(\frac{\sigma_y^2}{\sigma_x^2\sigma_y^2 - \xi^2}\right)\left(\ln x - \frac{\xi\ln y}{\sigma_y^2}\right)^2\right]\\ &= \frac{1}{x}\frac{1}{\sqrt{2\pi\sigma'^2}}\exp\left[-\frac{(\ln x - \mu')^2}{2\sigma'^2}\right],\quad (10)\end{aligned}$$

where the final expression defines $\mu'$ and $\sigma'^2$. Recall that when $x$ and $y$ are each Gaussian distributed, the conditional distribution of $x$ given $y$ is a shifted Gaussian. Equation (10) shows the analogous result for a lognormal distribution: when $x$ and $y$ are each distributed lognormally, the conditional distribution is a lognormal distribution (compare equations 10 and 5), whose associated underlying Gaussian is shifted. (The mean and variance of the underlying shifted Gaussian are $\mu'$ and $\sigma'^2$, respectively.) This means that if $x$ and $y$ are distributed lognormally, then an ensemble of realizations of $x$, with each realization subject to the (same) constraint $y$, will have a lognormal distribution.

Since the mean and the most probable values of a lognormal distribution are different, this means that the most probable value of $x$ given the constraint $y$, does not equal $\langle x|y\rangle$, the mean value of $x$ given $y$. This illustrates one important difference between lognormal and Gaussian distributions. When constructing constrained realizations of a lognormal field, it may be important to decide whether to construct realizations that resemble the mean realization, or whether to reconstruct the most probable realization. Below, expressions for both possibilities are derived.

The most probable value of $x$ given a value of $y$ is that value of $x$ for which

$$\frac{\mathrm{d}P_{\rm LN}(x|y)}{\mathrm{d}x} = 0,$$

so that

$$(x|y)_{\rm mp} = \exp\left[\frac{\xi\ln y}{\sigma_y^2} - \left(\frac{\sigma_x^2\sigma_y^2 - \xi^2}{\sigma_y^2}\right)\right] = e^{\mu' - \sigma'^2}, \quad (11)$$

where $\mu'$ and $\sigma'^2$ were defined in equation (10). On the other hand, the mean value

$$\begin{aligned}\langle x|y\rangle_{\rm LN} &= \int \frac{\mathrm{d}\ln x\, e^{\ln x}}{\sqrt{2\pi}}\sqrt{\frac{\sigma_y^2}{\sigma_x^2\sigma_y^2 - \xi^2}}\\ &\quad \times \exp\left[-\frac{1}{2}\left(\frac{\sigma_y^2}{\sigma_x^2\sigma_y^2 - \xi^2}\right)\left(\ln x - \frac{\xi\ln y}{\sigma_y^2}\right)^2\right]\\ &= \exp\left[\frac{\xi\ln y}{\sigma_y^2} + \left(\frac{\sigma_x^2\sigma_y^2 - \xi^2}{2\sigma_y^2}\right)\right]\\ &= e^{\mu' + (\sigma'^2/2)}.\quad (12)\end{aligned}$$

When $\xi \to 0$ (i.e., the limit where $\ln x$ and $\ln y$ are not correlated), then $(x|y)_{\rm mp} \to x_{\rm mp}$, and $\langle x|y\rangle_{\rm LN} \to \langle x\rangle_{\rm LN}$, as one would expect. In the other extreme, when $x$ and $y$ are completely correlated, so that $\xi = \sigma_x^2 = \sigma_y^2$, then $(x|y)_{\rm mp} = \langle x|y \rangle_{\rm LN} = y$, which is also expected. It is important to notice that, in general, equations (11) and (12) may be quite different from each other.

Next, it is necessary to compute the distribution function of the residuals from the mean realization (and also from the most-probable realization). To compute the distribution function of the residuals from the mean define $u = x - \langle x|y \rangle_{\rm LN}$ and compute $P(u|y)$. Then find the mean and the variance of this distribution. Recall that when $x$ and $y$ are Gaussian random variables, the mean and second moments of $u$ are independent of $y$. Here,

$$\langle u \rangle \equiv \int u\, P(u|y)\, {\rm d}u = 0 \qquad (13)$$

as expected. However,

$$\int u^2 P(u|y){\rm d}u = \langle x^2|y \rangle - \langle x|y \rangle^2 = e^{2\mu' + 2\sigma'^2} - e^{2\mu' + \sigma'^2}, \quad (14)$$

so that, as equations (10-12) show, the distribution of the residual field depends on the value of $y$.

Similarly, define the residuals from the most probable realization $v = x - (x|y)_{\rm mp}$ and compute the distribution function $P(v|y)$. Then

$$\langle v|y \rangle = \langle x|y \rangle - (x|y)_{\rm mp}, \qquad (15)$$

and

$$\langle v^2|y \rangle = \langle x^2|y \rangle - 2\,\langle x|y \rangle\, (x|y)_{\rm mp} + (x|y)_{\rm mp}^2, \qquad (16)$$

so that the distribution of $v$ also depends on $y$.

### 2.3 Generalization to random fields

Although the analysis above only considered the random variables $x$ and $y$ for which the mean of the underlying Gaussian variable was zero, it is clear that when considering a random field rather than a random variable, with some nonzero value for the underlying mean Gaussian field, nothing significant will be different. For this generalization to lognormal random fields the mean constrained realization is

$$\langle f(\boldsymbol{r})|\Gamma \rangle_{\rm LN} = \exp\left[\sum_{i,j} \xi_i(\boldsymbol{r})\xi_{ij}^{-1}\ln c_j \right.$$
$$\left. + \frac{\xi(0) - \xi_i(\boldsymbol{r})\xi_{ij}^{-1}\xi_j(\boldsymbol{r})}{2}\right], \qquad (17)$$

where the lognormal field is $f(\boldsymbol{r}) = f(r_1, \ldots, r_n)$, $\Gamma$ is the set of, say, $m$ constraints each of which specify, for example, that the value of the (lognormal) field at the $i$th point is $c_i$, $\xi_i(\boldsymbol{r})$ is the cross correlation between the underlying Gaussian field (that is associated with $f(\boldsymbol{r})$, the lognormal field) at $\boldsymbol{r}$ and the log of the $i$th constraint, $\xi(0)$ is the variance of the underlying Gaussian field, and $\xi_{ij}$ is the matrix that specifies the correlation between the log of the $i$th and $j$th constraints (it is the covariance matrix of the underlying Gaussian field that is associated with the lognormal constraint field). (This notation is similar to that of Hoffman & Ribak 1991.)

Similarly, the most probable realization is

$$(f(\boldsymbol{r})|\Gamma)_{\rm mp} = \exp\left[\sum_{i,j} \xi_i(\boldsymbol{r})\xi_{ij}^{-1}\ln c_j\right.$$

$$\left. -\xi(0) + \xi_i(\boldsymbol{r})\xi_{ij}^{-1}\xi_j(\boldsymbol{r})\right], \qquad (18)$$

and the variance of the residuals around the mean realization, $F(\boldsymbol{r}) = f(\boldsymbol{r}) - \langle f(\boldsymbol{r})|\Gamma \rangle_{\rm LN}$, is easily calculated from equation (17). As with the simpler random variable example considered earlier, this generalization to a random field shows that when the density field and the constraint field are both lognormal, then the distribution functions of the residuals from both the mean and the most probable realizations are a function of the constraints.

The results above suggest that, since the variance of the residuals is a function of the constraints, the Hoffman–Ribak technique may not be used to construct constrained realizations of lognormal fields. However, because a lognormal field is so easily related to its underlying Gaussian field (equation 4), there is a simple transformation that simplifies the problem considerably.

Equations (5) and (6) suggest defining a new $E$ field so that $\varepsilon \equiv \ln y$. Then the one-point pdf of $E$ is Gaussian and, as equation (6) shows, all the formalism developed for Gaussian random fields may be applied to the $E$ field. For example, the $n$-point pdfs of the $E$ field are all multivariate Gaussians. Now, equation (10) shows that the conditional distribution of a lognormal field, when the constraints are distributed lognormally, is just a lognormal whose underlying Gaussian is shifted. Therefore, for the $E$ field defined here, the conditional distribution for the $E$ field is just this shifted Gaussian (since, if the constraints are lognormal, then the log of the constraints is Gaussian). Therefore, by judicious choice of the constraint field, the Hoffman–Ribak algorithm may be used to construct constrained realizations of $E$.

In particular, if the constraint field essentially specifies the value of $\varepsilon$ (that is, it specifies the value of the log of the density field $y$) at a given set of points, then, if $Y$ is a lognormal, (so $E$ is Gaussian), the distribution of the constraints on $Y$ will be lognormal, so that the distribution of the constraints on $E$ will be Gaussian. Since the $E$ field and the constraint field are both Gaussian, and the conditional distribution is just a shifted Gaussian, the distribution of the residuals (from the mean of $E$) in any particular realization of the constrained field will be independent of the values of the constraints. So, the Hoffman–Ribak algorithm can be applied to construct constrained realizations of $E$. Then, simply taking the exponential of $E$ at every point gives a constrained realization of the lognormal field. The question is: how optimal is this procedure for constructing constrained realizations of the lognormal field, $Y$?

The following example illustrates this question. Assume that a constrained realization of the (Gaussian) $E$ field has been constructed (using the Hoffman–Ribak algorithm). This means that in between the points where the field is constrained to have certain values the field fluctuates around its mean realization value, with most of the fluctuations occurring within 'one standard deviation' or so. Since $E$ is Gaussian, this is perfectly acceptable as a realization of $E$. However, we are really interested in $Y = \exp E$. Taking the exponential of the $E$ field, at the points where the constraints have been specified, gives a $Y$ field that (by construction) has the correct values at these points. In between these points, however, as a result of the exponential trans-formation, the 'one standard deviation' departures from the mean $E$ realization become much larger departures from the mean $Y$ realization. It is not obvious that these fluctuations correctly sample the true (lognormal, cf. equation 10) distribution of fluctuations around the mean (lognormal) $Y$ realization. This section shows explicitly that this procedure (i.e., transforming the constrained realization of the underlying Gaussian field, $E$) does, indeed, give the optimal procedure for constructing constrained realizations of lognormal fields.

The mean and variance of the underlying constrained Gaussian variable $E$ are $\mu' = \xi \ln y / \sigma_y^2$ and $\sigma'^2 = (\sigma_x^2 \sigma_y^2 - \xi^2)/\sigma_y^2$, respectively. The mean and the most-probable values of the associated lognormal variable $Y$ are $e^{\mu'+(\sigma'^2/2)}$ and $e^{\mu'-\sigma'^2}$. Comparing with equations (11) and (12), we see that $(x|y)_{\rm mp}$ and $\langle x|y\rangle_{\rm LN}$ have precisely the same values as the mean and most probable values of the $Y$ field that is obtained by taking the exponential of the constrained $E$ field.

Furthermore, let $U(R)$ denote the variance of the residuals from the mean $Y = \exp E$ realization. That is, define

$$U(R) \equiv \left\langle \left(e^{\mu'+R} - e^{\mu'+(\sigma'^2/2)}\right)^2 \right\rangle. \qquad (19)$$

The first term in the angle brackets is $Y$, written explicitly as the exponential of the value of the variable $E$, which is written as the sum of its mean value, $\mu'$, plus a residual, $R$, from that mean. The second term is the mean value of the lognormal variable $Y$. Now, $E$ is Gaussian distributed, with mean $\mu'$ and variance $\sigma'^2$. This means that the residuals, $R$, from the mean $E$ realization have a Gaussian distribution with mean zero, and variance $\sigma'^2$, so that $p(R) = e^{-R^2/2\sigma'^2}/\sqrt{2\pi\sigma'^2}$. This, with equation (19), implies that

$$\begin{aligned} U(R) &= \left\langle e^{2\mu'+2R} - 2\, e^{\mu'+R} e^{\mu'+(\sigma'^2/2)} + e^{2\mu'+\sigma'^2} \right\rangle \\ &= e^{2\mu'} \left\{ \left[\int e^{2R} p(R)\,{\rm d}R\right] - \left[2 e^{\sigma'^2/2} \int e^R p(R)\,{\rm d}R\right] \right. \\ &\qquad \left. + \left[e^{\sigma'^2} \int p(R)\,{\rm d}R\right] \right\} \\ &= e^{2\mu'}\left(e^{2\sigma'^2} - 2\, e^{\sigma'^2/2}e^{\sigma'^2/2} + e^{\sigma'^2}\right) \\ &= e^{2\mu'+2\sigma'^2} - e^{2\mu'+\sigma'^2}, \end{aligned} \qquad (20)$$

where we have used the fact that

$$\begin{aligned} \left\langle e^{nR}\right\rangle &\equiv \int e^{nR} p(R)\,{\rm d}R = \int e^{nR}\frac{e^{-R^2/2\sigma'^2}}{\sqrt{2\pi\sigma'^2}}\,{\rm d}R \\ &= \int e^{n^2\sigma'^4/2\sigma'^2}\frac{e^{-(R-n\sigma'^2)^2/2\sigma'^2}}{\sqrt{2\pi\sigma'^2}}\,{\rm d}R = e^{\frac{n^2\sigma'^2}{2}}. \end{aligned} \qquad (21)$$

Comparing equations (20) and (14) shows that the variance of the residuals around the exponential of a constrained Gaussian variable, $U(R)$, is the same as the variance of the residuals around the mean of a constrained lognormal variable (equation 14).

Thus, this calculation shows that the ensemble of lognormal $Y$ fields that are obtained by taking exponentials of an ensemble of (appropriately) constrained realizations of $E$ fields, correctly samples the underlying constrained lognormal distribution of realizations. In other words, insofar as the Hoffman–Ribak algorithm provides an optimal technique for constructing constrained realizations of Gaussian random fields, the technique here, of constructing constrained realizations of the log of a lognormal field, and then taking the exponential of the constrained realization, is optimal for constructing constrained realizations of lognormal random fields. Thus, although the direct calculations of the previous subsection (and the complications involved in equations 14-16) suggested that the Hoffman–Ribak algorithm would not be optimal for constructing constrained realizations of a lognormal random field, with the transformation described here, the Hoffman–Ribak algorithm can, indeed, be used to construct optimal constrained realizations of a given lognormal density field.

To illustrate the method, Fig. 1 shows the steps involved in constructing a constrained realization of a lognormal distribution, given one constraint, say $C_1$, which specifies the value of the field at the origin. The top panel shows a cut through an unconstrained realization of a three-dimensional Gaussian random field that has power spectrum $P(k) \propto k^{-1} \exp[-(kR_s)^2]$. The Gaussian field is generated on a $32^3$ grid using FFT techniques, and is similar to the one presented in the top panel of Fig. 1 in Hoffman & Ribak (1991). The mean realization, given the value at the origin, $f(0)$, is shown by the dotted line. It is determined by $f(r) = f(0)\xi(r)/\xi(0)$, where $\xi(r) = \xi(|r-0|)$ is the autocorrelation function (i.e., $\xi(r)$ is the Fourier transform of the power spectrum), and $\xi(0)$ is the variance of the field. The dashed lines show the mean realization plus and minus one standard deviation, given the value at the origin, where the standard deviation $\sigma(r) = \sqrt{\xi(0) - [\xi(r)^2/\xi(0)]}$ (cf. Hoffman & Ribak 1991). At every point $r$, the residual is obtained by subtracting the value of the mean realization at $r$ (dotted line) from the actual value of the unconstrained field (solid line) at that position.

The dotted line in the middle panel shows the mean realization of the Gaussian when it is constrained to have a $2\xi(0)$ peak at the origin, (i.e. $f(0) = C_1 = 2\xi(0)$): $f(r) = C_1\,\xi(r)/\xi(0)$, and the dashed lines show the mean realization plus and minus one standard deviation. The solid line shows this mean plus the residual calculated from the previous panel; it is the constrained realization of the $2\xi(0)$ peak.

The solid line in the bottom panel shows the constrained realization of the corresponding lognormal field; it is obtained simply by taking the exponential of the solid line in the second panel. The dotted line shows the exponential of the mean of the constrained Gaussian realization. The dot-dashed line shows the ensemble mean of the constrained lognormal realizations, and at each point $r$, it is calculated using $f(r) = \exp[\mu'+(\sigma'^2/2)]$, where $\mu' = C_1\,\xi(r)/\xi(0)$, and $\sigma'^2 = \xi(0) - [\xi(r)^2/\xi(0)]$. The triple dot-dashed line shows the most-probable values of the constrained lognormal field, and is given by $f(r) = \exp[\mu' - \sigma'^2]$. These relations generalize those derived in equations (11-16).

The remainder of this section contains a brief digression on the philosophy of the approach of dealing with the log of a variable rather than its actual value. The next section gives a prescription for constructing a constrained realization of, for example, the *IRAS* catalogue.

Although the analysis above deals mainly with building constrained realizations of density fields, it can easily be

<ns>>6    *Ravi K. Sheth*</ns>

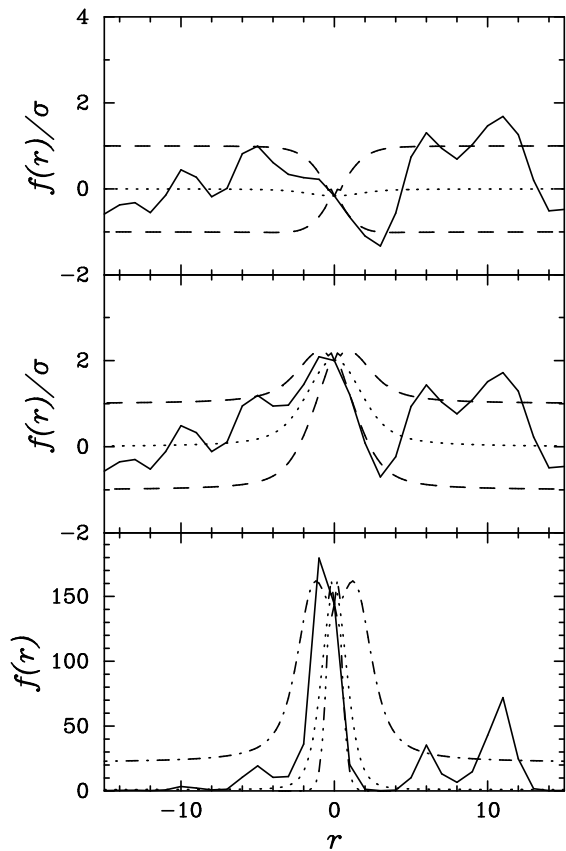

**Figure 1.** Constrained realizations of Gaussian and lognormal fields. *(Top)* Solid line shows a cut through an unconstrained 3-dimensional Gaussian random field that has power spectrum $P(k) \propto k^{-1} \exp[-(kR_s)^2]$. Dotted line shows the mean value of the realization given the value at the origin; dashed lines show mean plus and minus one standard deviation. *(Middle)* Solid curve shows the constrained Gaussian field constructed from the previous panel: it is constrained to have a peak that is two-standard deviations high at the origin. Dotted and dashed lines show the mean constrained realization, and the mean plus and minus one standard deviation, respectively. *(Bottom)* Solid curve shows the corresponding constrained lognormal field. Dotted curve shows the exponential of the mean field of the underlying Gaussian, i.e. the exponential of the dotted curve of the previous panel. Dot-dashed curve shows the mean constrained lognormal realization, and triple dot-dashed curve shows the most-probable constrained lognormal realization.

generalized. For example, it may be that one person (called A) is interested in building constrained realizations of some density field as measured by, say, its optical flux, whereas someone else (called B) is more interested in the field as measured by its optical magnitude, rather than by its optical flux. The latter is essentially the log of the former. If B were to build a constrained realization of the magnitude distribution, A could take the appropriate exponential transformation of B's realization to obtain an estimate of the distribution of flux. Provided the distribution of magnitudes is Gaussian, so that the distribution of flux is lognormal, not only will taking the exponential of the constrained realization of the magnitude distribution give a good indication of what the optical flux distribution looks like, but, as shown above, A's exponential transformation of B's constrained realization provides an optimal constrained realization of the distribution of flux.

Returning to the problem of constrained realizations of density fields, this suggests that if the true density distribution is really lognormal, then the convenient variables in which to work are not the values of the density field at a given point, but rather, the log of those values. If one is interested in just the log of the density field, and the density field is a lognormal, then the Hoffman–Ribak algorithm will give the 'optimal' constrained realization of that lognormal field. Furthermore, comparison of this constrained realization with the log of, say, the *IRAS* distribution, is a fair way of comparing the constrained realization to the data. Perhaps the way to put this is as follows: If the density field, as measured by its *IRAS* flux is lognormal, then, when applied to the log of the flux, the Hoffman–Ribak algorithm will give the optimal constrained realization of the magnitude distribution. Taking the exponential of the constrained realization will give an optimal realization of the flux field.

## 3  CONSTRUCTING THE CONSTRAINED FIELD

The best way to construct a constrained realization of a lognormal field that is similar to, say, the *IRAS* distribution, even on small highly nonlinear scales, is as follows. Construct a constrained realization of the $E$ field using the Hoffman–Ribak algorithm. Next, obtain the $Y$ field by taking the exponential of the $E$ field. Then use the prescription of Coles and Jones (1991, Section 8.3) to convert the density field $Y$ into a point distribution. One could even arrange the conversion so that the number of points in the constrained realization is the same as the number of galaxies in the *IRAS* catalogue. Finally, smooth the point distribution with a filter of choice to look at the density distribution on any scale of interest. Compare with the result when the *IRAS* catalogue itself is smoothed with that same filter.

The all important first step is to construct a constrained realization of the $E$ field. This requires knowledge of the covariance function of the log of the *IRAS* distribution; once this covariance function is known, the following steps are straightforward. This covariance function is obtained as follows. The two-point correlation function of the random variable $y$ is

$$\xi(r) \equiv \frac{\langle (y_1 - \langle y \rangle)(y_2 - \langle y \rangle) \rangle}{\langle y \rangle^2}, \qquad (22)$$

where $r = |\mathbf{r}_1 - \mathbf{r}_2|$, $y_1$ and $y_2$ are the values of the field at $\mathbf{r}_1$ and $\mathbf{r}_2$, and $\langle y \rangle$ is the mean value of $y$. If $y$ is distributed lognormally, then its mean can be calculated similarly to equation (12); it is

$$\langle y \rangle = \exp\left(\mu + \frac{\sigma^2}{2}\right), \qquad (23)$$

where $\mu$ and $\sigma^2$ denote the mean and variance of the underlying Gaussian. For this lognormal, $\xi(r)$ is easily calculated from the two-point pdf:

$$1 + \xi(r) \;=\; \frac{1}{\langle y \rangle^2} \int\!\!\int y_1 y_2 \, f(y_1, y_2) \, dy_1 \, dy_2$$

$$= \frac{1}{\langle y \rangle^2} \exp\left(\frac{\sigma_1^2}{2} + \frac{\sigma_2^2}{2} + 2\mu + \rho\right), \quad (24)$$

where $\sigma_1^2 = M_{11} = M_{22} = \sigma_2^2$, and $\rho = M_{12} = \langle (\ln y_1 - \mu)(\ln y_2 - \mu) \rangle$. This final term shows that the covariance function, $\rho$, is related to the correlation function of the underlying Gaussian (the correlation function is $\rho/\mu^2$). Writing $\Xi(r)$ for the covariance function of the Gaussian means that $\rho = \Xi(r)$, so equations (23) and (24) show that

$$1 + \xi(r) = \exp[\Xi(r)]. \quad (25)$$

(e.g. Vanmarcke 1983; Coles & Jones 1991).

As Coles & Jones note, this exact relation resembles the Politzer & Wise (1984) approximation used in biased clustering theory. Politzer & Wise were trying to calculate the correlation function for high-level regions of a Gaussian random field. They found that, to a good approximation, the correlation function, $\xi(r)$, of high-level regions of a Gaussian field is related to the correlation function, $\Xi(r)$, of the underlying Gaussian by $1 + \xi(r) = \exp[\Xi(r)]$. Since galaxies are thought to be biased traces of the matter distribution, this provides another (albeit heuristic) justification for considering a lognormal field as a good model for the galaxy distribution—the correlation function of the high density regions of a Gaussian field (the dark matter distribution) looks just like the correlation function of a lognormal field (the observed galaxy distribution).

Equation (25) shows that, although it has not been measured directly, the covariance function of the log of the *IRAS* distribution, $\Xi(r)$, can be obtained from the *IRAS* correlation function itself. So, to construct a constrained realization of the $E$ field only requires a reliable measurement of the *IRAS* correlation function, $\xi(r)$, and of $\mu \equiv \langle \ln y \rangle$.

To make constrained realizations of the projected distribution, $\xi(r)$ should be replaced by $\omega(\theta)$. The *IRAS* angular correlation function can be obtained in a number of ways: (1) directly measuring $\omega(\theta)$, (2) via Limber's equation from the spatial correlation function measured by, e.g., Saunders, Rowan-Robinson & Lawrence (1992), (3) from the 3-D power spectrum (Fisher et al. 1993), or (4) from the angular power spectrum. Although Fourier transforming the *IRAS* power spectrum is the natural choice when making constrained realizations of the large scale distribution (e.g. Lahav et al. 1994), since this paper is concerned with smaller scale features (hence the move from the Gaussian to the lognormal), it is more natural to use the correlation function itself, since, on these small scales, it is a better estimator of the correlations. Most measurements of the correlation function [using either (1) or (2) above] give

$$\omega(\theta) = \left(\frac{\theta_0}{\theta}\right)^\gamma, \quad (26)$$

with $\gamma \approx 1.6$ and $\theta_0 \approx 0.11°$ out to about $6°$ after which it drops quickly to zero.

With equation (26) for the correlation function, and with the relations given earlier that relate the covariance function of the underlying Gaussian to the correlation function of the lognormal, it should be relatively straightforward to construct constrained realizations of lognormal random fields that are similar to the *IRAS* distribution.

## 4 MINIMUM VARIANCE RECONSTRUCTIONS OF LOGNORMAL RANDOM FIELDS

If a quantity whose true value is $s$ is measured to have the value $y = s + n$, where $n$ is some noise on the measurement, then the Wiener filtered minimum variance linear reconstruction of $s$, given the measured $y$, is

$$s_{\text{wf}} = \mathbf{S}[\mathbf{S} + \mathbf{N}]^{-1} y = \mathbf{F} y \quad (27)$$

where $\mathbf{S} \equiv \langle ss^{\text{T}} \rangle$, and $\mathbf{N} \equiv \langle nn^{\text{T}} \rangle$. This relation holds whatever the true distribution of $s$, whatever the measured distribution of $y$, and whatever the distribution of the noise, $n$: there is no requirement that any of these fields be Gaussian (cf. Rybicki & Press 1992).

Lahav et al. (1994) show that if $s$ and $n$ are Gaussian random fields, so $y$ is also a Gaussian random field, then the minimum variance reconstruction, $s_{\text{mv}}$, is the same as the Wiener filtered reconstruction, which is the same as the most probable reconstruction defined using the constrained realizations formalism, and recall that, for Gaussian random fields, this most probable realization is also the mean realization. Since, in general, the mean and the most probable fields are different for a lognormal, it is interesting to compare the Wiener Filter reconstructed field, subject to the condition that the true field is lognormal, with the mean and most probable constrained lognormal fields.

We consider two cases. First, we consider a lognormal signal $s$, in which the noise on the underlying Gaussian field is assumed to be Gaussian and additive, so that $\log y = \log s + \log n$, where $y$ is the measured value of $s$, and $\log s$ and $\log n$ are both Gaussian. (This means that $y = sn$; the noise is multiplicative, not additive.) Then it is straightforward to apply the minimum variance reconstruction to this underlying Gaussian field (rather than to the lognormal field directly). Consider the field that is obtained by taking the exponential of the minimum variance reconstruction of the underlying Gaussian field:

$$s_{\text{est}} = e^{(\log s)_{\text{wf}}} = \exp\left(\mathbf{F} \log y\right). \quad (28)$$

Now, as mentioned above, the Wiener filter, minimum variance estimator of a Gaussian field is also the mean of the underlying constrained field (e.g., Lahav et al. 1994). So, in the notation of the previous sections, this reconstructed field is simply $e^{\mu'}$, where $\mu'$ denotes the mean value of the constrained field. However, recall that the most probable value is $e^{\mu' - \sigma'^2}$, whereas the mean value is $e^{\mu' + (\sigma'^2/2)}$ (compare equations 11 and 12). Thus, the estimator given in equation (28), i.e., the exponential of the Wiener filtered value of the underlying Gaussian field, has the attractive property of always being between the most probable and the mean constrained realizations.

The second case we consider is when $s$ is lognormal and $n$ is Gaussian. This case of Gaussian noise, whatever the underlying signal, may be useful for reconstruction analyses of small-scale features in, e.g., the microwave background. A simple example that uses the random variable, rather than the random field will illustrate the results. Let $y = s + n$, where $s$ is a lognormal distributed signal, and $n$ is Gaussian noise. Then $\log y = \log(s + n) = \log s + \log(1 + n/s)$. In the limit where $s$ is large relative to the noise, this simplifies to $\log y \approx \log s + (n/s)$. Since $n$ is Gaussian distributed, for sufficiently large $s$, we can treat $n/s$ as Gaussian distributed also. Therefore, in this approximation, $\log y$ is the sum of two Gaussian variables, so it too is Gaussian. So, this case is similar to that discussed above. This means that $\exp{(\log s)_{\rm wf}} = \exp{(F \log y)}$, is a good estimator of the underlying signal, since it is always between the most probable and the mean reconstructions. However, when the noise is not small, which is the more general and useful case, obtaining a useful estimator of the underlying signal is more complicated.

To obtain an expression for the most probable value of the signal given the measured value $y$, consider $P(s|y) = P(s,y)/P(y) = P(s)P(y|s)/P(y)$, where the distribution $P(s)$ is lognormal, and $P(y|s) = P(s+n|s) = P(n)$ is Gaussian. Since $P(y) = \int P(s)P(n)\,{\rm d}s$, with $n = y - s$, is merely a normalizing factor, there is no need to compute it explicitly. Below, we will compare the minimum variance (Wiener filtered) estimate, $s_{\rm wf}$, of the underlying signal given the measured value $y$, with the most probable value of $P(s|y)$. This most probable value, $s_{\rm mp}$, is given by requiring that ${\rm d}P(s|y)/{\rm d}s = 0$. This requirement means that

$$\begin{aligned}
\frac{{\rm d}P(s|y)}{{\rm d}s} &= \frac{1}{P(y)}\frac{\rm d}{{\rm d}s}P(s)P(n{=}y{-}s) \\
&= \frac{P(n{=}y{-}s)}{P(y)}\frac{{\rm d}P(s)}{{\rm d}s} + \frac{P(s)}{P(y)}\frac{{\rm d}P(n{=}y{-}s)}{{\rm d}s} \\
&= \frac{P(n)P(s)}{P(y)}\left(-\frac{1}{s} - \frac{\ln s}{s\sigma_s^2}\right) + \frac{P(s)P(n)}{P(y)}\frac{(y-s)}{\sigma_n^2} \\
&= \frac{P(n)P(s)}{P(y)}\left(\frac{y-s}{\sigma_n^2} - \frac{1}{s} - \frac{\ln s}{s\sigma_s^2}\right) = 0, \qquad (29)
\end{aligned}$$

where we have assumed that $s$ is distributed lognormally in such a way that the mean of the underlying Gaussian, $\mu_s = \langle \ln s \rangle = 0$ and $\langle (\ln s)^2 \rangle = \sigma_s^2$, and that the noise is Gaussian distributed around zero with variance $\sigma_n^2$. This implies that $s_{\rm mp}$ and $y$ satisfy the nonlinear relation

$$y = s_{\rm mp} + \frac{\sigma_n^2}{s_{\rm mp}} + \frac{\sigma_n^2}{\sigma_s^2}\frac{\ln s_{\rm mp}}{s_{\rm mp}}. \qquad (30)$$

On the other hand, the relation between the Weiner filter estimate, $s_{\rm wf}$, and the measured value, $y$, is linear. Namely,

$$s_{\rm wf} = \frac{\langle (s - \langle s \rangle)^2 \rangle}{\langle (s - \langle s \rangle)^2 \rangle + \langle n^2 \rangle}\,y = \frac{e^{2\sigma_s^2} - e^{\sigma_s^2}}{e^{2\sigma_s^2} - e^{\sigma_s^2} + \sigma_n^2}\,y. \qquad (31)$$

Thus, the relation between $s_{\rm mp}$ and $s_{\rm wf}$ is complicated and, in general, the Wiener filter estimate differs from the most probable one: $s_{\rm wf} \ne s_{\rm mp}$. The difference between the two quantities depends on the variance, and hence on the power spectra of the signal and the noise. When the noise is vanishingly small, so that $\sigma_n^2 \to 0$, then both $s_{\rm mp}$ and $s_{\rm wf} \to y$, as they should. When the variance of the noise is much larger than that of the signal, then $s_{\rm mp} \to e^{-\sigma_s^2}$ as expected (cf. equation 11). In contrast, the Wiener filter estimate is 0, in this limit. Finally, note that if $s$ is distributed lognormally, then $s$ must always be positive. Notice that the estimator $s_{\rm mp}$ is positive even when $y$ is negative. In contrast, $s_{\rm wf}$ has the same sign as the measured value, $y$, so it can be negative.

Since the relation between $s_{\rm mp}$ and $s_{\rm wf}$ is nonlinear, the accuracy of the Wiener filter estimate of the underlying signal is most easily illustrated numerically. Before doing so, it is useful to extend equation (30) to the case of a measured random field $\boldsymbol{y} = \boldsymbol{s} + \boldsymbol{n}$, where $\boldsymbol{s}$ is an underlying lognormal signal and $\boldsymbol{n}$ is the noise associated with the measurement, and is assumed to be Gaussian. Then, the most probable estimator of the underlying signal $\boldsymbol{s}$ satisfies the relation

$$\boldsymbol{s}^{-1} + {\rm diag}\left(\boldsymbol{s}^{-1}\right)\,\boldsymbol{\Xi}^{-1}\,\ln \boldsymbol{s} + \mathbf{N}^{-1}\boldsymbol{s} = \mathbf{N}^{-1}\,\boldsymbol{y}, \qquad (32)$$

where $\mathbf{N}$ and $\boldsymbol{\Xi}$ are the covariance matrices of the noise and of the log of the signal, respectively. Compared to the Wiener filter estimator (equation 27), this equation is much more difficult to solve.

In the absence of correlations, both $\boldsymbol{\Xi}$ and $\mathbf{N}$ are diagonal, so that the most probable estimator of $\boldsymbol{s}$ at a given point in the field is independent of the values of $\boldsymbol{s}$ or $\boldsymbol{n}$ at other points in the field. Thus, in the absence of correlations, equation (32) reduces to equation (30) at each point. Similarly, the Wiener filter estimate (equation 27) reduces to equation (31) at each point. Of course, this is consistent with equation (25) which shows that if the underlying Gaussian field is uncorrelated, so that its covariance matrix is diagonal, then the covariance matrix of the associated lognormal field is also diagonal.

Fig. 2 shows a numerical example of the difference between the most probable and the Wiener Filter estimators of the underlying signal for the case when the noise, $\boldsymbol{n}$ is Gaussian and the signal, $\boldsymbol{s}$, is lognormal, and both have diagonal covariance matrices. The light solid line in the top panel shows $\boldsymbol{y} = \boldsymbol{s} + \boldsymbol{n}$, whereas the heavy solid line shows the actual value of the signal, $\boldsymbol{s}$. Both $\boldsymbol{s}$ and $\boldsymbol{y}$ are plotted in units of the root mean square value of the (Gaussian) noise. In the Figure, $s_{\rm rms} = 5.23$ at each point, which corresponds to an underlying Gaussian with zero mean and dispersion $\sigma_s^2 = 1.75$. This value was chosen because it gives approximately the same value of the variance of the galaxy density field on the scales on which the lognormal may be good approximation. For convenience, the (Gaussian) noise has zero mean and dispersion equal to that of the signal, $s_{\rm rms}^2$.

Since the covariance matrices are diagonal, the Wiener filter estimator is $\boldsymbol{s}_{\rm wf} = \boldsymbol{y}/2$ at every point. The light solid line in the bottom panel shows this Wiener Filter estimator. The heavy solid line in the bottom panel shows $\boldsymbol{s}_{\rm mp}$; at each point it is the largest root of equation (30) that is less than the measured value $\boldsymbol{y}$ at that point. For large values of $\boldsymbol{y}$, $s_{\rm mp}$ is nearly the same as $\boldsymbol{y}$, which means that it, in the Figure, is a factor of two larger than the Wiener Filter estimator (large values of $\boldsymbol{y}$ are all those that are greater than, say, $3\sigma_n$). For those values of $\boldsymbol{y}$ that are intermediate (say, between $\sigma_n$ and $3\sigma_n$), the most probable estimator is comparable to the Wiener filter estimator. For those values of $\boldsymbol{y}$ that are less than $\sigma_n$, $s_{\rm mp}$ is approximately $e^{-\sigma_s^2} \sim 0$. Finally, note that whereas $\boldsymbol{s}_{\rm wf}$ is often negative, $\boldsymbol{s}_{\rm mp}$ is always positive.

Fig. 2 shows that it is possible for $\boldsymbol{s}_{\rm wf}$ to differ substantially from $\boldsymbol{s}_{\rm mp}$. As stated above, the exact difference between $\boldsymbol{s}_{\rm wf}$ and $\boldsymbol{s}_{\rm mp}$ will depend on the details of the covariance matrices of the noise and the signal (equations 27 and 32). Thus, this Section shows explicitly that if the underlying signal is non-Gaussian, then the Wiener Filter estimator of the underlying signal, given a noisy measurement of that signal, may not be very close to the most probable value of that signal, given the same measurement.

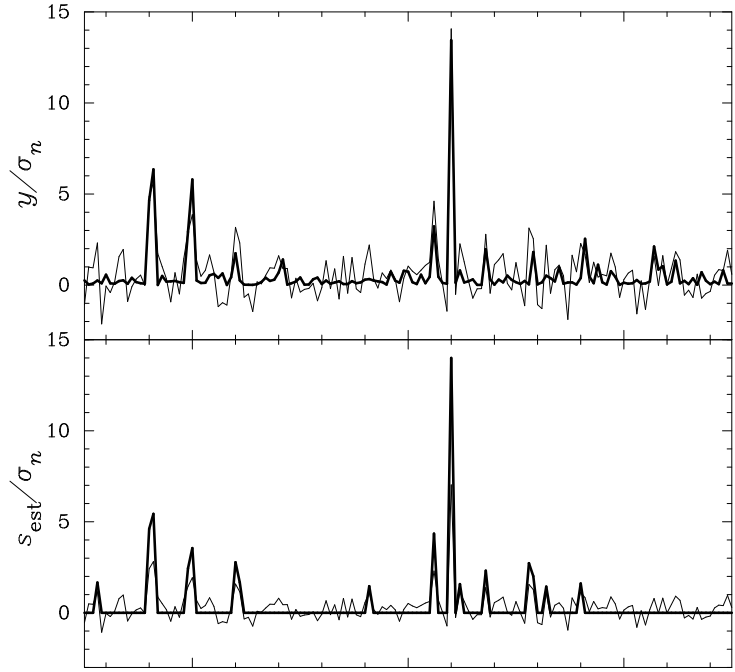

**Figure 2.** Example of the difference between the most probable (heavy solid line, bottom panel) and the Wiener filter (light solid line, bottom panel) estimators of the underlying signal for the case when the noise, $n$ is Gaussian and the signal, $s$, is lognormal, and both have diagonal covariance matrices. Light solid line in the top panel shows $y = s + n$. Heavy solid line in top panel shows the actual value of the signal, $s$.

## 5 DISCUSSION AND EXTENSIONS

It is straightforward to extend the technique developed in this paper for constructing constrained realizations of lognormal random fields to other non-Gaussian fields that can be obtained by transforming an underlying Gaussian field. Here we consider two other examples, the generalized Chi-squared random field with $n$ degrees of freedom (hereafter $\chi_n^2$), and the generalized Rayleigh distribution; these random fields generalize some of the distributions considered by Coles & Barrow (1987).

The characteristic function of a $p$-variate $\chi_n^2$-field is

$$G_n(\boldsymbol{t}) = \left(\det||\mathbf{I} - i2\mathbf{MT}||\right)^{-n/2}, \qquad (33)$$

where $G_n(\boldsymbol{t})$ denotes $G_n(t_1, t_2, \ldots, t_p)$, $\mathbf{I}$ is the identity matrix, $\mathbf{T}$ is the diagonal matrix with diagonal elements $t_1, t_2, \ldots, t_p$, and $\mathbf{M}$ is the covariance matrix of the underlying Gaussian field (e.g. Lukacs 1977). Although, in general, the corresponding probability density $f_n(\boldsymbol{z}) = f_n(z_1, z_2, \ldots, z_p)$ with

$$z_i = \sum_{j=1}^{n} x_j^2, \qquad (34)$$

where the $x_j$s are independent normal variates is complicated, a calculation of the case when $p = 2$ and the $x_j$s have the same unit variance and normal distribution around mean zero, will serve to illustrate our purpose. Then,

$$\frac{f_n(z_1, z_2)}{f_n(z_1)f_n(z_2)} = \exp\left(\frac{-\rho^2(z_1+z_2)}{(1-\rho^2)}\right) \frac{\Gamma(\frac{n}{2})}{(1-\rho^2)}$$
$$\times \left(\rho^2 z_1 z_2\right)^{-\frac{(n-2)}{4}} I_{\frac{(n-2)}{2}}\left(\frac{2\rho\sqrt{z_1 z_2}}{1-\rho^2}\right), (35)$$

where

$$f_n(z)\,\mathrm{d}z = \frac{z^{(n-2)/2}}{\Gamma(\frac{n}{2})}\,e^{-z}\,\mathrm{d}z \qquad (36)$$

provided $z > 0$, and $f_n(z) = 0$ otherwise, $\Gamma(z)$ is the Gamma function, $I_\alpha(z)$ is the modified Bessel function, and $\rho = \langle x_1 x_2 \rangle$ is the covariance function of the underlying Gaussian variables (e.g. Vere-Jones 1967).

From these expressions, and using the series expansion of the modified Bessel function, the conditional distribution, $f_n(z_2|z_1) = f_n(z_1, z_2)/f_n(z_1)$, is easily computed; it is

$$f_n(z_2|z_1)\,\mathrm{d}z_2 = \frac{z^{(n-2)/2} e^{(z+\lambda)/2}}{2^{n/2}}$$
$$\times \sum_{r=0}^{\infty} \frac{\lambda^k z^k}{2^{2k}\,k!\,\Gamma(\frac{n}{2}+k)} \frac{2\,\mathrm{d}z_2}{(1-\rho^2)}, (37)$$

where $z = 2z_2/(1-\rho^2)$, $\lambda = 2\rho^2 z_1/(1-\rho^2)$. Equation (37) for the conditional distribution of $z_2$ given $z_1$ is a non-central $\chi_n^2$ distribution (e.g. Miller 1964, p. 56; Stuart & Ord 1991, Sections 23.4, 23.6), so it can be understood as arising from a shifted Gaussian distribution. More generally, if $\mu_1$ and $\mu_2$, and $\sigma_{x_1}^2$ and $\sigma_{x_2}^2$, are, respectively, the means and variances of the original Gaussians, the mean and variance of the underlying (shifted) Gaussian distribution are $\mu' = \rho\sqrt{(z_1/n)}/\sigma_{x_1}^2$, and $\sigma'^2 = (\sigma_{x_1}^2 \sigma_{x_2}^2 - \rho^2)/\sigma_{x_1}^2$, respectively. By analogy with the lognormal example discussed earlier (compare equation 10), it is clear that taking the appropriate transformation of the (appropriately) constrained Gaussian field will give constrained realizations of $\chi_n^2$-fields that have the correct ensemble properties.

Recently Bunn et al. (1994) used constrained realizations of Gaussian random fields to assess the statistical sig-nificance of hot and cold spots in maps of the microwave background radiation as seen by the *COBE* DMR. As Luo (1995) and Coulson et al. (1994) note, the data presently available admit the possibility that on small ($\sim 0.5°$) angular scales, fluctuations in the radiation may have a non-Gaussian spatial distribution. Luo (1994) argues that $\chi_n^2$-fields provide a family of random fields that range from the highly non-Gaussian (low-$n$) to the nearly Gaussian (large $n$), and so they are able to model a large range of non-Gaussian distributions. In addition, temperature fluctuations from topological and non-topological defects in the framework of the $O(N)\sigma$-model are thought to be well described by a $\chi_n^2$-field (Turok & Spergel 1991). Both these considerations justify a continued interest in $\chi_n^2$-fields. They also suggest that this generalization of the Hoffman–Ribak algorithm to $\chi_n^2$-fields may be useful for assessing the statistical significance of structure in the microwave background on small angular scales.

Additionally, as Sheth et al. (1994) show, the discrete Negative Binomial distribution provides a good fit to the distribution function of galaxy counts-in-cells. The Negative Binomial can be understood as arising from a Poisson sampling process [exactly the same Poisson process discussed by Layzer (1956) and Peebles (1980), and analogous to that used by Coles & Jones (1991) to construct a discrete number count model from a (continuous) lognormal density field], applied to an underlying Gamma distribution which has the form shown in equation (36) (e.g. Stuart & Ord 1991, Section 16). So, it appears that Poisson sampling of an appropriately constrained realization of a $\chi_n^2$-field, should produce constrained realizations of the Negative Binomial distribution that are similar to the observed galaxy distribution.

Coles & Barrow (1987) studied the lognormal, the $\chi_n^2$, and two other non-Gaussian distributions that can be obtained by transforming a Gaussian, as models of the distribution of extreme hot spots in the microwave background. For completeness, we now study a generalization of their Rayleigh and Maxwell distributions. Expressions for the generalized Rayleigh distribution, $g_n(r_1, r_2, \ldots, r_p)$, where

$$r_i = \sqrt{\sum_{j=1}^{n} x_j^2}, \tag{38}$$

and where all the $x_j$s are independent normal variates are given, e.g., in Miller (1964, ch. 2). Here, we simply consider the case when $p = 2$ [the case when $p = 1$ and $n = 2$ is the Rayleigh distribution studied by Coles & Barrow (1987), and they termed the $p = 1$ with $n = 3$ case the Maxwell distribution]. Then, when the $x_j$s all have zero means and unit variance,

$$g_n(r) = \frac{2 r^{n-1} e^{-r^2/2}}{2^{n/2}\, \Gamma(\frac{n}{2})}, \tag{39}$$

and

$$g_n(r_1, r_2) = \frac{(r_1 r_2)^{n/2}}{(2\rho)^{(n-2)/2}} \frac{1}{(1-\rho^2)\Gamma(\frac{n}{2})}$$
$$\times \exp\left(\frac{-(r_1^2 + r_2^2)}{2(1-\rho^2)}\right) I_{\frac{(n-2)}{2}}\left(\frac{r_1 r_2 \rho}{1-\rho^2}\right), \tag{40}$$

where $\rho$ and $\Gamma(r)$ are defined as for the $\chi_n^2$ case. With these two expressions it is straightforward to show that $g(r_2|r_1)$ is a Rayleigh distribution whose underlying Gaussian variables are shifted [see Miller (1964) for the expression that describes a generalized Rayleigh distribution with non-zero mean and non-unit variance]. In general, the mean and variance of the shifted Gaussians are $\mu' = \rho r_1/\sqrt{n}\sigma_{x_1}^2$ and $\sigma'^2 = (\sigma_{x_1}^2 \sigma_{x_2}^2 - \rho^2)/\sigma_{x_1}^2$, respectively. As shown explicitly for the lognormal case, it is clear that taking the appropriate transformation of the appropriately constrained Gaussian field will give constrained realizations of generalized Rayleigh fields that have the correct ensemble properties.

All these results strongly support the speculation that, with the appropriate transformations of the constraint field, the Hoffman–Ribak algorithm is optimal for constructing ensembles of constrained realizations of all random fields that are obtained by transforming Gaussian fields. A sketch of the proof is given in the Appendix.

## 6 CONCLUSIONS

Section 2.3 showed that, with appropriate modifications, the Hoffman–Ribak algorithm can be used to construct an ensemble of constrained realizations of lognormal density fields that has the correct ensemble properties. Fig. 1 showed the steps involved in constructing such a constrained lognormal field. Section 3 provided an algorithm that can produce constrained realizations of lognormal fields that can be compared directly with, e.g., the spatial distribution of galaxies in the *IRAS* catalogue.

Since the mean and the most probable values of any lognormal field are different, the accuracy of minimum variance reconstruction estimates of lognormal random fields is less than when reconstructing Gaussian random fields (for which the mean and the most probable are the same). Section 4 provided two models for the relation between a measurement, the noise and an underlying lognormal signal. When the signal is lognormal and the noise is both lognormal and multiplicative (rather than additive), then the Wiener filter reconstruction provides a robust estimate of the underlying signal, since it always lies between the mean and the most probable values. However, when the signal, $s$, is lognormal and the noise, $n$, is Gaussian and additive, (so the measured quantity is $y = s + n$), then the accuracy of the Wiener filter reconstruction depends on the power spectrum of the signal and of the noise. Equations (27) and (32) showed that, in general, the most probable reconstructed value is different from the Wiener filter estimate. Fig. 2 provided a simple numerical example of the way in which these two estimators differ. It showed that the most probable reconstruction differs significantly from the Wiener filter estimate at those points where the measured values are greater than $3\sigma_n$, where $\sigma_n$ is the root mean square of the noise. Therefore, when the underlying signal is lognormal, then the accuracy of the Wiener filter reconstruction will differ for different data sets.

Finally, Section 5 showed that the Hoffman–Ribak algorithm can be extended to construct ensembles having the correct ensemble properties of constrained realizations of all random fields that are obtained by transforming Gaussian fields. The proof was sketched in the Appendix.

## ACKNOWLEDGMENTS

I thank the Marshall Aid Commemoration Commission for a Marshall Scholarship, the Institute of Astronomy and Jesus College, Cambridge, for financial support, Marc Davis and the Berkeley Astronomy Department for their hospitality, Xiaochun Luo for discussions about $\chi_n^2$ random fields, Ofer Lahav and Yehuda Hoffman for encouragement during the early parts of this work, and Saleem Zaroubi for encouragement at the end.

## APPENDIX A: CONDITIONED NON-GAUSSIAN RANDOM FIELDS

Following, e.g., Miller (1964), let $X = \{x_1, x_2, \ldots, x_n\}$ be an $n$-dimensional Gaussian vector with the positive definite, symmetric, covariance matrix $\mathbf{M}$. So

$$f(X)\,dX = \frac{dX}{(2\pi)^{n/2}\sqrt{|\mathbf{M}|}} \exp\left(-\frac{1}{2} X^T \mathbf{M}^{-1} X\right), \quad (A1)$$

where $f(X)\,dX$ denotes the probability that the vector lies between $X$ and $X + dX$, and $|\mathbf{M}|$ denotes the determinant of $\mathbf{M}$ (compare equation 2). The covariance matrix, $\mathbf{M}$, may be partitioned into a $p \times p$ matrix, $\mathbf{C}_1$, in the top left corner, an $(n-p) \times (n-p)$ matrix, $\mathbf{C}_2$, in the bottom right corner, a $p \times (n-p)$ matrix, $\mathbf{B}$ in the top right corner, and the transpose of $\mathbf{B}$, $\mathbf{B}^T$, an $(n-p) \times p$ matrix, in the bottom left corner. Since $\mathbf{M}$ is symmetric, $\mathbf{C}_1 = \mathbf{C}_1^T$ and $\mathbf{C}_2 = \mathbf{C}_2^T$.

Further, since $\mathbf{M}$ is nonsingular, its inverse, $\mathbf{M}^{-1}$, exists, and it too can be partitioned.

Let $\mathbf{Q}^{-1}$, $\mathbf{P}^{-1}$, $\mathbf{R}$, and $\mathbf{R}^T$ denote the corresponding partitions of $\mathbf{M}^{-1}$. Since $\mathbf{M}$ is symmetric, $\mathbf{Q} = \mathbf{Q}^T$ and $\mathbf{P} = \mathbf{P}^T$. Then $\mathbf{P}\mathbf{R}^T = -\mathbf{B}^T \mathbf{C}_1^{-1}$ and $\mathbf{Q}^{-1} = \mathbf{C}_1^{-1} + \mathbf{R}\mathbf{P}\mathbf{R}^T$ are Shur's identities, and $|\mathbf{P}| = |\mathbf{M}||\mathbf{C}_1|^{-1}$, where the vertical bars denote the determinants of the matrices, is a special case of Jacobi's theorem (cf. Miller 1964, Section 1.3).

Finally, let $X_1 = \{x_1, x_2, \ldots, x_p\}$, with $1 \le p \le n$, and let $X_2 = \{x_{p+1}, x_{p+2}, \ldots, x_n\}$. So, $X_1$ is the $p$-dimensional vector comprised of the first $p$ elements of $X$, and $X_2$ is the $(n-p)$-dimensional vector comprised of the last $(n-p)$ elements of $X$. This means that

$$\begin{aligned} X^T \mathbf{M}^{-1} X &= X_1^T \mathbf{Q}^{-1} X_1 + X_2^T \mathbf{R}^T X_1 \\ &\quad + X_1^T \mathbf{R} X_2 + X_2^T \mathbf{P}^{-1} X_2 \\ &= (X_2 + \mathbf{P}\mathbf{R}^T X_1)^T \mathbf{P}^{-1} (X_2 + \mathbf{P}\mathbf{R}^T X_1) \\ &\quad + X_1^T (\mathbf{Q}^{-1} - \mathbf{R}\mathbf{P}\mathbf{R}^T) X_1, \end{aligned} \quad (A2)$$

where the second expression follows by adding and subtracting $X_1^T \mathbf{R}\mathbf{P}\mathbf{R}^T X_1$ to complete the square in the $X_2$ variables. However, the second of Shur's identities shows that

$$X_1^T (\mathbf{Q}^{-1} - \mathbf{R}\mathbf{P}\mathbf{R}^T) X_1 = X_1^T \mathbf{C}_1^{-1} X_1. \quad (A3)$$

So, setting $V = -\mathbf{P}\mathbf{R}^T X_1$ means that

$$X^T \mathbf{M}^{-1} X = (X_2 - V)^T \mathbf{P}^{-1} (X_2 - V) + X_1^T \mathbf{C}_1^{-1} X_1. \quad (A4)$$

Notice that $V = -\mathbf{P}\mathbf{R}^T X_1 = \mathbf{B}^T \mathbf{C}_1^{-1} X_1$, where the second equality follows from the first of Shur's identities (compare with the terms in equation 17), is independent of $X_2$.

Now, it is clear that

$$f(X_1)\,dX_1 = \frac{dX_1}{(2\pi)^{p/2}\sqrt{|\mathbf{C}_1|}} \exp\left(-\frac{1}{2} X_1^T \mathbf{C}_1^{-1} X_1\right), \quad (A5)$$

and recall that $f(X) = f(X_1, X_2)$. So, the conditional distribution, $f(X)/f(X_1)$, is given by:

$$\begin{aligned} f(X_2|X_1)\,dX_2 &= \frac{f(X_1, X_2)\,dX_1\,dX_2}{f(X_1)\,dX_1} \\ &= \frac{dX_2}{(2\pi)^{(n-p)/2}} \sqrt{\frac{|\mathbf{C}_1|}{|\mathbf{M}|}} \times \exp \\ &\quad \left(-\frac{1}{2}(X_2 - V)^T \mathbf{P}^{-1} (X_2 - V)\right). \end{aligned} \quad (A6)$$

Since $|\mathbf{P}| = |\mathbf{M}||\mathbf{C}_1|^{-1}$ (Jacobi's theorem) and since $V$ is independent of $X_2$, this shows that if $f(X_1, X_2)$ is Gaussian distributed, then the conditional distribution, $f(X_2|X_1)$, is a shifted Gaussian.

Now consider the vector $Y$, where $Y = g(X)$, and where $X$ is a Gaussian vector with covariance matrix $\mathbf{M}$. That is, the vector $Y$ is some transformation of the Gaussian vector $X$. For example, if $Y = \exp X = \{e^{x_1}, e^{x_2}, \ldots, e^{x_n}\}$, then $dX = d\log Y = \prod_i dy_i/y_i$, and the distribution of $Y$ is lognormal (compare equation 6). Similarly, if $Y = \{x_1^2, x_2^2, \ldots, x_n^2\}$, then $Y$ is a $\chi^2$-(with one degree of freedom)-vector, and $dX = d\sqrt{Y} = \prod dy_i/2\sqrt{y_i}$ (compare equation 37). Clearly, if $Y = g(X)$, where $X$ is a Gaussian vector, and if $Y$ is partitioned into $Y_1$ and $Y_2$, then the conditional distribution, $f(Y_2|Y_1) = f(Y_1, Y_2)/f(Y_1)$, will be given by the appropriate transformation of the underlying Gaussian that has been shifted by $V = -\mathbf{P}\mathbf{R}^T X_1 = \mathbf{B}^T \mathbf{C}_1^{-1} g^{-1}(Y_1)$. As a result, with the appropriate modifications described in the text, the Hoffman–Ribak algorithm may be used to construct constrained realizations of $Y$ that have the correct ensemble properties.

Of course, this assumes that $g^{-1}(Y_1)$ is single valued. For the lognormal distribution this is satisfied. For the Chi-squared distribution, however, both the positive and the negative square roots are valid solutions. In practise, this will not be a problem if the mean of the underlying Gaussian field is known to be positive, and if the (transformed) constraints are unlikely to be negative. This will almost always be the case if, for example, the constraints are chosen to correspond to sufficiently high density peaks of the Chi-squared field.